\documentclass{revtex4}  %% REVTeX 4.0
\usepackage[latin1]{inputenc}
\usepackage{amsmath}
\usepackage{amsfonts}
\usepackage{amssymb}
\usepackage{graphicx,color}
\usepackage[%
  colorlinks=true,
  urlcolor=blue,
  linkcolor=blue,
  citecolor=blue
]{hyperref}

\begin{document}
%%%%%%%%%%%%%%%%%% title page information %%%%%%%%%%%%%%%%%%
\title{Imaging spatio-temporal Hong-Ou-Mandel interference of biphoton state of extremely high Schmidt number}

\author{Fabrice Devaux$^{1}$, Alexis Mosset$^1$, Paul-Antoine Moreau$^2$, and Eric Lantz$^1$}
\affiliation{$^1$ Institut FEMTO-ST, D\'epartement d'Optique P. M. Duffieux, UMR 6174 CNRS \\ Universit\'e Bourgogne Franche-Comt\'e, 15b Avenue des Montboucons, 25030 Besan\c{c}on, France\\
$^2$ School of Physics and Astronomy, University of Glasgow, G12 8QQ, UK\\}
\date{\today}
%\email[Corresponding author:]{fabrice.devaux@univ-fcomte.fr}

\begin{abstract}
We report the experimental observation of a spatio-temporal Hong-Ou-Mandel (HOM) interference of biphoton state of extremely high Schmidt number. Two-photon interference of  1500 spatial modes and a total of more than $3\times10^6$ spatio-temporal modes is evidenced by measuring momentum spatial coincidences, without any prior selection of the photons in time and space coincidence, between the pixels of the far-field images of two strongly multimode spontaneous parametric down conversion (SPDC) beams propagating through a HOM interferometer. The outgoing SPDC beams are recorded on two separate detectors arrays operating in the photon-counting regime. The properties of HOM interference are investigated both in the time and space domains. We show that the two-photon interference exhibits temporal and two-dimensional spatial HOM dips with visibilities of 60\% and widths in good agreement with the spatio-temporal coherence properties of the biphoton state. Moreover, we demonstrate that maxima of momentum spatial coincidences are evidenced within each image, in correspondence with these dips.            
\end{abstract}

%\pacs{42.65.Lm,03.67.Bg}

\maketitle
\section{Introduction}
Spatial entanglement of photon pairs in images offers new opportunities to develop protocols for communication and parallel treatment of quantum informations of potentially very high dimensionality. Although entangled photon pairs of high Schmidt number are easily produced by SPDC, the manipulation and the detection of images with quantum features is tricky. Fortunately, detector arrays with high sensitivity such as Electron Multiplying Coupled-Charge Device (EMCCD), intensified charge coupled device (iCCD) or single photon avalanche diode (SPAD) array \cite{lubin_quantum_2019} are now widely used for quantum imaging experiments \cite{moreau_imaging_2019} like demonstration of Einstein-Podolsky-Rosen (EPR) paradox in twin images \cite{moreau_realization_2012,moreau_einstein-podolsky-rosen_2014,lantz_einstein-podolsky-rosen_2015}, imaging of high-dimensional spatial entanglement \cite{edgar_imaging_2012}, ghost imaging \cite{morris_imaging_2015,denis_temporal_2017}, quantum adaptive optics \cite{defienne_adaptive_2018}, quantum holography \cite{devaux_quantum_2019}, sub-shot-noise imaging \cite{brida_experimental_2010,toninelli_sub-shot-noise_2017} and quantum imaging with undetected photons \cite{lemos_quantum_2014}.

Among the whole experiments using entangled pairs of photons, the famous experiment of two-photon interference known now as Hong-Ou-Mandel (HOM) interference \cite{hong_measurement_1987}, is probably one of the most fascinating. This groundbreaking experiment paved the way for a multitude of experiments showing the richness of the quantum properties of light and their application to original communication protocols \cite{simon_quantum_2017}, to quantum teleportation \cite{bouwmeester_experimental_1997} and in the context of quantum information and computation, e.g. in linear optical quantum computing \cite{kok_linear_2007} and more particularly in boson sampling \cite{gard_introduction_2015}. Most of these experiments and protocols used the coherence time properties of the biphoton state and the measurements are performed by means of bucket detectors and coincidence counters gated in time. HOM interference is obtained if the two involved photons are indistinguishable, whatever their origin, meaning that extremely dissimilar light sources \cite{deng_quantum_2019} can be used if the corresponding modes are thoroughly tailored. On the other hand, genuine multi-mode HOM interference implies entanglement, as quoted by Lee et al \cite{lee_spatial_2006}, and SPDC remains the simplest way to produce entangled photon pairs of high dimensionality. Recently, Jachura et al. \cite{jachura_shot-by-shot_2015} extended the applications of the camera systems to the observation of HOM interference with an intensified scientific complementary metal-oxide-semiconductor (sCMOS) camera, showing a maximum of coincidences on the same region of interest (ROI) of the camera in conditions corresponding to a minimum of the dip between separate ROIs of the camera. Nevertheless, the input photons were spatially filtered by traversing a single-mode fiber to ensure an unique input spatial mode for each photon. It is possible to imprint a phase profile on one of the photon and to realize its hologram, as demonstrated by the same group \cite{chrapkiewicz_hologram_2016}. Also in this experiment, an only spatial mode per input port is involved, even if shaped. In the experiments reported in Refs \cite{ou_further_1989,kim_spatial_2006}, it was shown that a tilt between the spatially monomode input beams results in coincidence fringes, that were detected by scanning a bucket detector. On the other hand, two experiments demonstrated multi-mode HOM interference. First, Walborn et al. \cite{walborn_multimode_2003} showed that a HOM dip can be transformed in a HOM peak by using either an antisymmetric pump beam or an anti-symmetric polarization vector of the entangled photon pair. Second, Lee et al. employed bucket detectors and time coincidence to characterize HOM interference from SPDC limited by an aperture \cite{lee_spatial_2006} or whose orbital angular momentum (OAM) has been modified by an image rotator \cite{di_lorenzo_pires_measurement_2010}, resulting in a maximum of about 40 spatial modes or 20 OAM modes.

Recently, using numerical simulations with realistic parameters, we have shown how two cameras can be used to detect two-dimensional (2D) spatial coincidences of biphoton state of high dimensionality and to evidence HOM interferences between the two outgoing images \cite{devaux_stochastic_2019}. Thank to this numerical model, we have demonstrated that such a HOM interferometer allows the characterization of the temporal as well as the 2D spatial coherence properties of entangled photons pairs.

 In  this paper, we report the first experimental observation of a fully spatio-temporal HOM interference of biphoton state of high Schmidt number. Two-photon interferences are evidenced by measuring the 2D momentum spatial coincidences between the pixels of far-field images of twin SPDC beams propagating through the HOM interferometer without any prior selection of the photons in time and space coincidence. The images are recorded onto two EMCCDs operating in the photon-counting regime \cite{lantz_multi-imaging_2008}. The use of EMCCDs allows the detection of all  photons of the images and the measurement of spatial coincidences on the whole set of photons.The properties of HOM interference are investigated both in the time and space domains, as it was proposed in \cite{devaux_stochastic_2019}. Given the critical role played by two photon HOM interferences in most quantum information and quantum technology schemes, our demonstration that HOM interference can be obtained by manipulating a very high dimensional entangled state paves the way to very high dimensional quantum information schemes using space and time variables.           

\section{Experimental setup and method}
\begin{figure}
\centering
\includegraphics[width=15cm]{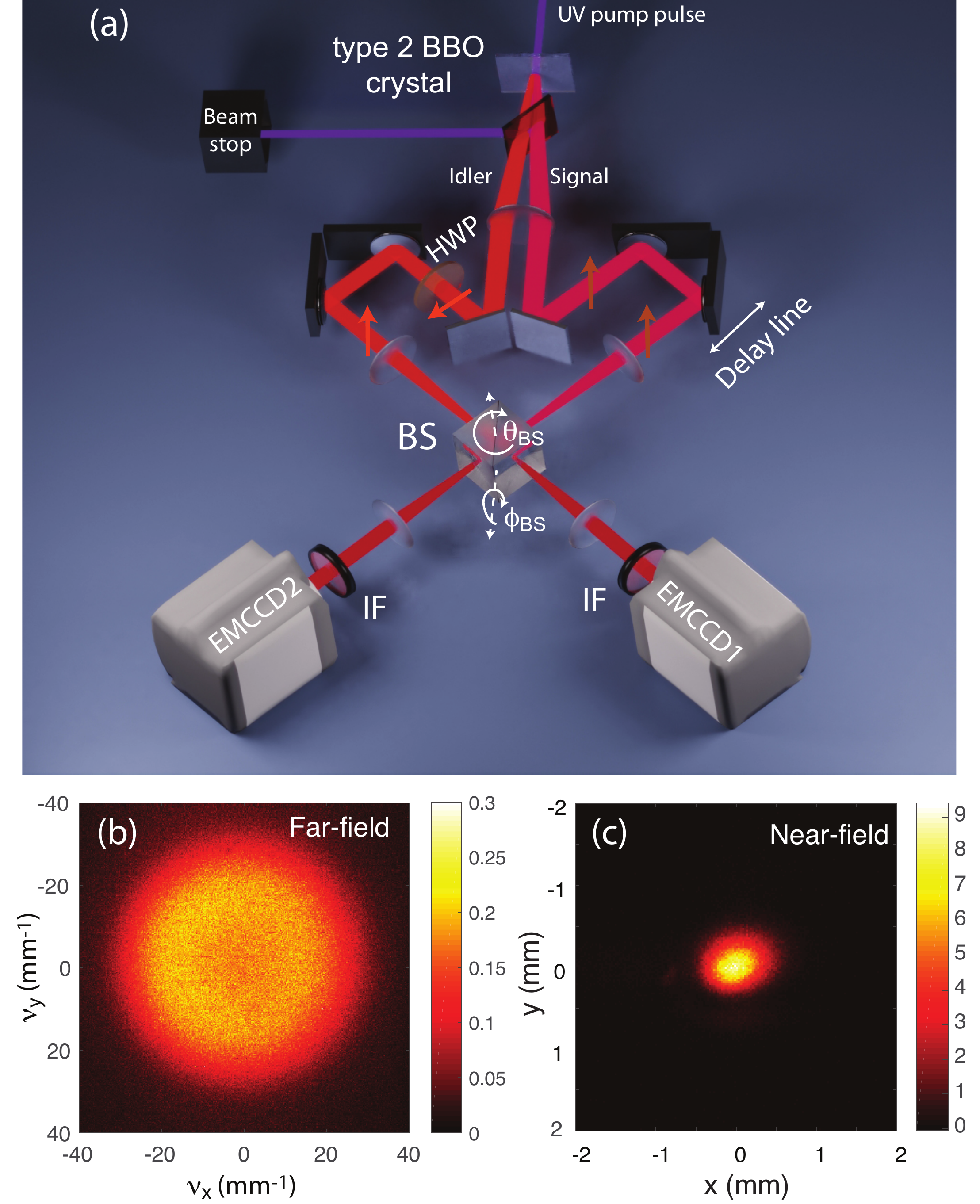}
\caption{(a) Experimental setup. (b) and (c) average images in photon number of the far-field and the near-field of a single SPDC beam, respectively.}\label{setup}
\end{figure}
Fig. \ref{setup}a shows the experimental setup that closely corresponds to the setup modelized in \cite{devaux_stochastic_2019}. Strongly multimode twin SPDC beams (i.e. biphoton state of high Schmidt number) are generated in a noncolinear type-II geometry in a $0.8\,mm$ long $\beta$-barium borate ($\beta$-BBO) crystal pumped at $355\,nm$. The pump pulses are provided by a passively Q-switched Nd:YAG laser ($\Delta t_{pump}=660\,ps$ FWHM pulse duration, 8 $mW$ mean power and 4 $kHz$ repetition rate). Because of the noncolinear interaction, the twin beams are separated and propagate through the two input ports of the HOM interferometer up to a beamsplitter (BS: $R$=50\%, $T$=40\%, losses 10\%). In both arms, two identical 1.5 magnification telescopes form the near-field images of the BBO crystal inside the BS. Because of the geometry of the interferometer \cite{devaux_stochastic_2019} where SPDC beams propagate in the horizontal plane, the reflected beams experience a left-right symmetry with respect to the transmitted beams in the near-field as well as in the far-field. Before the BS, the polarization state of the idler beam is controlled with a half-wave plate (HWP), in order to measure spatial correlations when the polarizations states of the twin photons are horizontal-vertical ($HV$) or vertical-vertical ($VV$). Then, the far-field of the two outgoing images is formed with $2f$ imaging systems on two separate EMCCDs (ANDOR iXon Ultra 897), used in photon-counting regime \cite{lantz_multi-imaging_2008}. Before detection, the photons pairs emitted around the degeneracy are selected by narrow-band interference filters (IF) centered at $709\,nm$ ($\Delta\lambda_{IF}\backsimeq 5\,nm$, FWHM bandwidth). Figures \ref{setup}b and \ref{setup}c show typical far-field and near-field average images of the SPDC beams through the HOM interferometer.  The axes of the far-field image are graduated in spatial frequency coordinates $\nu_{x,y}$, which are related to momentum coordinates by $q_{x,y}=2\pi\nu_{x,y}$. 

The control, by a delay line, of the time delay $\delta t$ between the input ports of the HOM interferometer gives access to the coherence length. Meanwhile, the control of a 2D transverse spatial frequency shift between the transmitted and reflected beams at the output ports, by the rotations $\delta\theta_{BS}$ and $\delta\phi_{BS}$ of the BS, gives access to the two transverse coherence widths of the biphoton state. From the far-field and near-field images of the SPDC beams (figures \ref{setup}b and \ref{setup}c), the time duration of the pump pulse and the bandwidth of the IF, we estimate the standard deviations of the SPDC beams in the spatial and temporal domains as:
\begin{eqnarray}\label{eq1}
\left\{\begin{array}{c}
\sigma^{SPDC}_x\simeq\sigma^{pump}_x\simeq 0.35\,mm\\
\sigma^{SPDC}_y\simeq\sigma^{pump}_y\simeq 0.37\,mm\\
\sigma^{SPDC}_{\nu_{x}}\simeq 34 \,mm^{-1}\\
\sigma^{SPDC}_{\nu_{y}}\simeq 34 \,mm^{-1}\\
\sigma^{SPDC}_{t}\simeq\sigma^{pump}_{t}\simeq 400\,ps\\
\sigma^{SPDC}_{\nu_{t}}\simeq\sigma^{IF}_{\nu_{t}}\simeq 1.8\,THz\\ 
\end{array}\right.
\end{eqnarray}
In the context of entangled twin photons \cite{law_analysis_2004}, the Schmidt number is given along the dimension $i$ (where $i=x,\,y,\,t$) by $K_i=\frac{1}{2} \left(\sigma^{pump}_{i}2\pi\sigma^{SPDC}_{\nu_i}+\frac{1}{\sigma^{pump}_{i}2\pi\sigma^{SPDC}_{\nu_i}} \right)$. From the experimental parameters (Eq. \ref{eq1}), we have estimated the Schmidt numbers for each dimension:  
\begin{eqnarray}\label{eq2}
\left\{\begin{array}{c}
K_x\simeq 37\\
K_y\simeq 40\\
K_t\simeq 2.3\times 10^3\\
\end{array}\right.
\end{eqnarray}
Finally, the full space-time dimensionality of the biphoton state can be estimated as $K_xK_yK_t\simeq 3.4\times10^{6}$, which confirms the extremely high dimensionality of the twin photons involved in the HOM interference.
  
\begin{figure}
\centering
\includegraphics[width=15cm]{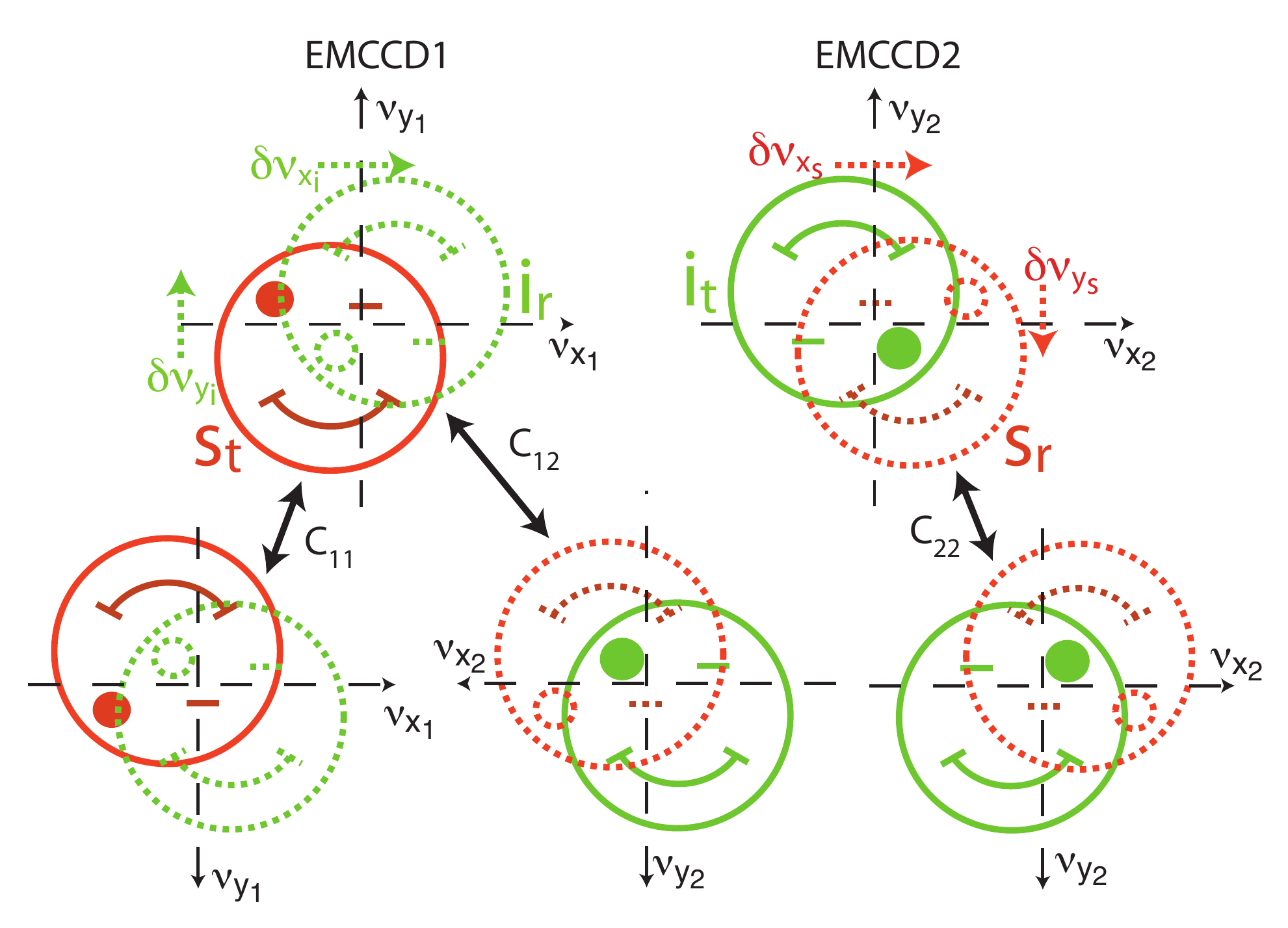}
\caption{(a) and (b) diagrams illustrating, on both cameras, the relative positions of the far-field patterns of the signal and idler beams, represented in red and green respectively, when they are either transmitted ($s_t$, $i_t$) or reflected ($s_r$, $i_r$) as a function of the shifts $\delta\nu_{x_i}=\delta\nu_{x_s}$ and $\delta\nu_{y_i}=-\delta\nu_{y_s}$ induced by horizontal and vertical tilts of the BS, respectively. The solid and dotted line circles correspond to the transmitted beams and to the reflected beams, respectively. The origins of the axes are centered on the barycenters of the images. (c) and (e) up-down flip applied to images (a) and (b). (d) left-right and up-down flips applied to image (b). Spatial correlations can be observed within images symmetrically to the horizontal axes ($C_{11}$ and $C_{22}$) and between images symmetrically to the images barycenters ($C_{12}$).}\label{IntraInter}
\end{figure}

Figures \ref{IntraInter}a and \ref{IntraInter}b illustrate schematically the relative positions and orientations of the far-field patterns of the SPDC beams on both cameras, when they are either transmitted or reflected by the BS. Because of the geometry of the HOM interferometer \cite{devaux_stochastic_2019}, horizontal and vertical tilts of the BS induce momenta shifts $\boldsymbol{\delta q}$ of the reflected beams which are related to 2D spatial frequency shifts $\delta\nu_{x_i}=\delta\nu_{x_s}=2f \delta\theta_{BS}$ and $\delta\nu_{y_i}=-\delta\nu_{y_s}=2f \delta\phi_{BS}$, where $f$ is the focal length of the last lenses before the EMCCDs. Using detectors arrays, the momentum spatial correlations between the photons of a pair can be measured between images for transmitted-transmitted ($tt$) and reflected-reflected ($rr$) twin photons and also within single images for reflected-transmitted ($rt$) and transmitted-reflected ($tr$) twin photons. These different kinds of spatial correlations can be observed by comparing figures \ref{IntraInter}a and \ref{IntraInter}c and figures \ref{IntraInter}b and \ref{IntraInter}e for correlations within single images and by comparing figures \ref{IntraInter}a and \ref{IntraInter}d for correlations between images.

First, let us consider the case where the transmitted and reflected beams are perfectly superimposed on both cameras ($\boldsymbol{\delta q}=\boldsymbol{0}$), perfectly synchronized ($\delta t=0$) and parallely polarized (VV configuration). For both reflected ($rr$) and transmitted beams ($tt$), momentum correlations are found between pixels of the two cameras corresponding to opposite transverse momenta coordinates $\boldsymbol{q_1}$ and $\boldsymbol{q_2}=-\boldsymbol{q_1}+\boldsymbol{\Delta q}$, where $\boldsymbol{\Delta q}$ denotes the 2D momentum correlation uncertainty. By calculating the normalized cross-correlation between the image 1 and the up-down and left-right symmetric image 2, we will obtain the spatial distribution of the momentum correlations as a function of the 2D momentum uncertainty : $C_{12}(\boldsymbol{\Delta q})$ . For the $rt$ and $tr$ beams, momentum correlations are found within each image between the pixels symmetric with respect to the horizontal axis, corresponding to transverse momenta coordinates $(q_{x1},\,q_{y1})$ and $(q_{x1}+\Delta q_x,\,-q_{y1}+\Delta q_y)$ for camera 1 and $(q_{x2},\,q_{y2})$ and $(q_{x2}+\Delta q_x,\,-q_{y2}+\Delta q_y)$ for camera 2. For each image, measurements are performed by calculating the normalized cross-correlation between the upper half and the up-down symmetric lower half parts of single images. In that case, we do not perform the correlation calculation on the whole image because the correlations between twin photons detected on the same camera would be counted twice. Then, we obtain the spatial distributions of the momentum correlations within each image as a function of the 2D momentum uncertainty: $C_{11}(\boldsymbol{\Delta q})$ and $C_{22}(\boldsymbol{\Delta q})$. In all cases, the correlation distributions have the shape of 2D gaussian functions with standard deviations related to the spatial dimensionality of the biphoton state. Now, let us consider the case where a time delay and a momentum shift are imposed. In that case, using the formalism proposed in \cite{devaux_stochastic_2019} and \cite{hong_measurement_1987}, we can establish the relations that give the spatial distributions of momentum correlations as a function of $\delta t$ and $\boldsymbol{\delta q}$:

\begin{eqnarray}\label{eq3}
\left\{\begin{array}{c}
C_{12}\left(\boldsymbol{\Delta q};\delta t,\boldsymbol{\delta q}\right)=\\
\left[R^2C_0\left(\boldsymbol{\Delta q +\delta q_x}\right)+T^2C_0\left(\boldsymbol{\Delta q -\delta q_x}\right)\right]\times\\
\left(1-\dfrac{2RT}{R^2 + T^2}e^{-\frac{\delta q^2_{x}}{\sigma_{q}^2}}e^{-\frac{\delta q^2_{y}}{\sigma_{SPDC}^2}}e^{-\frac{\delta t^2}{\sigma^2_{t}}}\right)\\
C_{11}\left(\boldsymbol{\Delta q};\delta t,\boldsymbol{\delta q}\right)+C_{22}\left(\boldsymbol{\Delta q};\delta t,\boldsymbol{\delta q}\right)=\\
RT\left[C_0\left(\boldsymbol{\Delta q +\delta q_x}\right)+C_0\left(\boldsymbol{\Delta q -\delta q_x}\right)\right]\times\\
\left(1+e^{-\frac{\delta q^2_{x}}{\sigma_{q}^2}}e^{-\frac{\delta q^2_{y}}{\sigma_{SPDC}^2}}e^{-\frac{\delta t^2}{\sigma^2_{t}}}\right)\\
\end{array}\right.
\end{eqnarray}

$C_0(\boldsymbol{\Delta q})$ is the spatial distribution of momentum correlations measured between twin images when the BS is removed.
According to these equations, we should measure two correlation peaks centered along the horizontal axis at a distance related to the horizontal spatial shift and with amplitudes related to the spatial and temporal shifts. As it was demonstrated in \cite{devaux_stochastic_2019}, the standard deviations of the spatial HOM dip depend of the coherence width of the biphoton wave-packet along the horizontal dimension $\sigma_{q}$ and of the phase matching bandwidth $\sigma_{SPDC}$ along the vertical dimension. $\sigma_t$ is the standard deviation of the temporal HOM dip. 
A 2D space integration of the equations \ref{eq3}, normalized by a 2D space integration of $C_0(\boldsymbol{\Delta q})$, leads to the relative spatial correlations as a function of the momentum shift and the time delay as follows:
 \begin{eqnarray}\label{eq4}
 \left\{\begin{array}{c}
 R_{12}(\delta t,\boldsymbol{\delta q})=R^2+T^2-2RTe^{-\frac{\delta q^2_{x}}{\sigma_{q}^2}}e^{-\frac{\delta q^2_{y}}{\sigma_{SPDC}^2}}e^{-\frac{\delta t^2}{\sigma^2_{t}}}\\
 R_{11}(\delta t,\boldsymbol{\delta q})+R_{22}(\delta t,\boldsymbol{\delta q})=\\
 2RT\times\left(1+e^{-\frac{\delta q^2_{x}}{\sigma_{q}^2}}e^{-\frac{\delta q^2_{y}}{\sigma_{SPDC}^2}}e^{-\frac{\delta t^2}{\sigma^2_{t}}}\right)\\
 \end{array}\right.
 \end{eqnarray}
 Eq. \ref{eq4} clearly shows that when twin photons are indistinguishable, the drop in spatial correlations between the two images (HOM dip) is accompanied by an increase in spatial correlations within the individual images (HOM maximum). Using the definition of the HOM dip visibility \cite{lee_spatial_2006}: $V_{12}=\frac{R^{max}_{12}-R^{min}_{12}}{R^{max}_{12}}=\frac{2RT}{R^2+T^2}$ and defining the visibility of the HOM maximum as  $V_{11+22}=\frac{\left(R_{11}+R_{22}\right)_{max}-\left(R_{11}+R_{22}\right)_{min}}{\left(R_{11}+R_{22}\right)_{max}}=\frac{1}{2}$, visibilities of 98\% of the HOM dip and 50\% of the HOM maximum are expected, considering the measured values of R and T ($R=$50\% and $T=$40\%).   
\section{Experimental results}  
\subsection{Temporal coherence measurements}
\begin{figure}
  \centering
  \includegraphics[width=15cm]{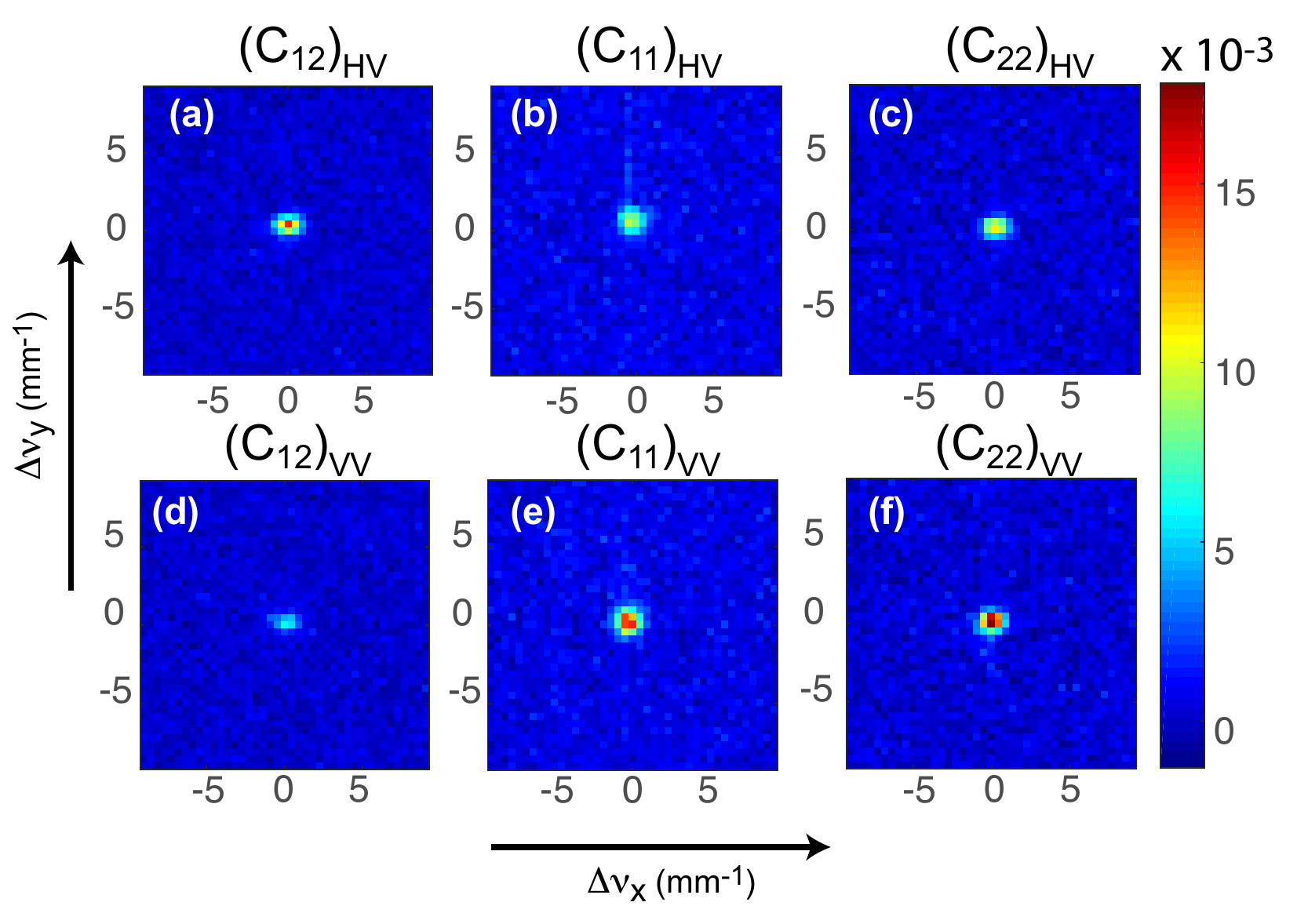}
  \caption{Average normalized spatial momentum correlations distributions obtained when correlations are measured between image pairs ($C_{12}$) and within single images ($C_{11}$, $C_{22}$) when the optical paths difference between the two arms of the interferometer is null. $HV$ and $VV$ indices are related to the polarization states of the twin photons.}\label{CORRspatialDiptemporel}
  \end{figure}

 The first experiment consists in measuring the temporal coherence of the biphoton state. To this end, the signal and the idler beams are spatially superimposed as precisely as possible in the near-field and in the far-field domains. Then, spatial momentum correlations are measured, between pairs of images as well as within single images, as a function of the optical path delay between the two arms of the HOM interferometer and as a function of the polarization states $HV$ and $VV$ of the twin photons. Fig. \ref{CORRspatialDiptemporel} shows the spatial momentum correlations distributions obtained when they are measured between image pairs ($C_{12}$) and within single images ($C_{11}$, $C_{22}$) and when the time delay between the two arms of the interferometer is null. These results are averaged over 500 pairs of images. Figures \ref{CORRspatialDiptemporel}a to \ref{CORRspatialDiptemporel}c are related to the $HV$ polarization states and figures \ref{CORRspatialDiptemporel}d to \ref{CORRspatialDiptemporel}f to the $VV$ polarization states. From these figures, we can observe that the three correlation peaks have the same gaussian-like shapes with standard deviations  $\sigma_{\nu_x}=0.8\,\pm 0.2\,mm^{-1}$ and $\sigma_{\nu_y}=0.6\,\pm 0.2\,mm^{-1}$. It means that spatial momentum correlations are measured with the same precision between images or within single images. Then, we can roughly estimate the spatial dimensionality of the biphoton state as $\dfrac{\sigma^{SPDC}_{\nu_x}\sigma^{SPDC}_{\nu_y}}{\sigma_{\nu_x}\sigma_{\nu_x}}=\dfrac{34^2}{0.8\times 0.6}\backsimeq 2400$. This result is of the same order of magnitude as the product of the estimated $K_x$ and $K_y$ Schmidt numbers (Eq. \ref{eq2}) and confirms the high spatial dimensionality of the biphoton state. By integrating the correlation peaks, we have estimated the ratio of the events corresponding to the detection of photons by pairs between images and within single images. When twin photons are cross-polarized ($HV$), these ratios are $13\pm 2 \%$ in $C_{12}$ and $7\pm 1 \%$ in $C_{11}$ and $C_{22}$. It means that more or less half of the spatial momentum coincidences are recorded between the two images and the other half is equally distributed within the single images. For the $VV$ polarization states, the ratio in $C_{12}$ decreases to $5\pm 1 \%$ and increases up to $10\pm 1 \%$ in $C_{11}$ and $8\pm 1 \%$ in $C_{22}$. Consequently, the decreasing of the ratio of spatial coincidences between images and its increasing within single images is clearly the demonstration that a HOM interference occurs for the biphoton state of high Schmidt number. For $HV$ and $VV$ configurations, the sum of the correlation ratios is somewhat smaller than the 25\% ratio measured between twin images without the BS, probably because of the losses of the BS.
        
\begin{figure}
 \centering
 \includegraphics[width=15cm]{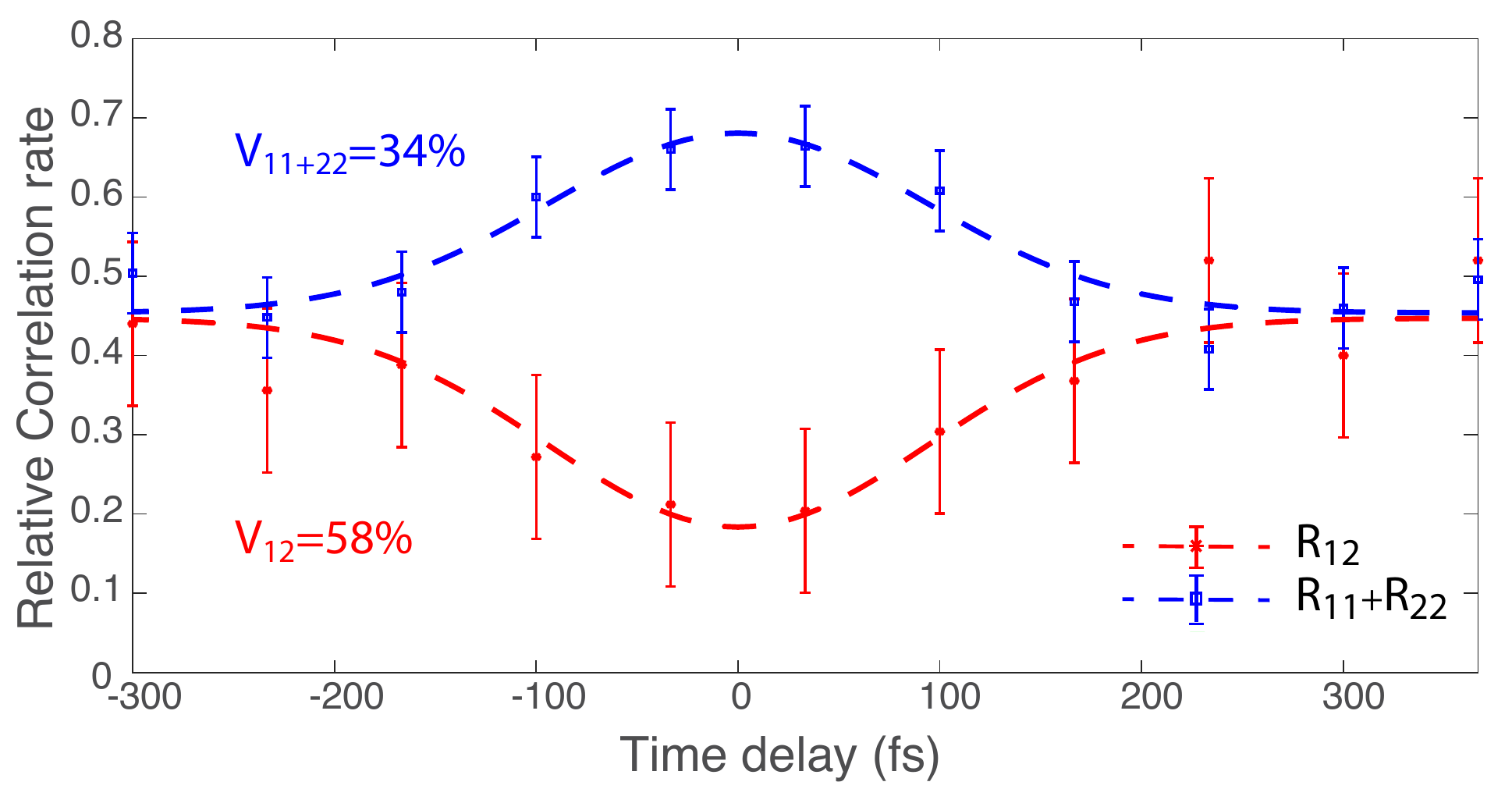}
 \caption{Relative spatial correlations ratios as a function of the time delay between the two arms of the HOM interferometer when the polarization states of the twin beams are vertical-vertical (VV). The red stars and blue squares correspond to correlations between twin images and within single images, respectively. The dotted curves correspond to nonlinear fits of the experimental data.These curves exhibit a HOM dip and a HOM maximum with visibilities of 58\% and 34\%, respectively.}\label{Diptemporel}
 \end{figure}

 In order to estimate the coherence length of the biphoton state we have measured the variation of the relative ratios of total photon-pairs detection events in images as a function of the time delay for $VV$ polarisation states. These relative ratios (Eq. \ref{eq4}) correspond to the number of detected photon-pairs measured between twin images as well as within single images, divided by the number of detected photon-pairs without the BS. Results are depicted in Fig. \ref{Diptemporel}. As expected, while a HOM dip is clearly exhibited for correlations between images, a HOM maximum is observed for correlations within single images. From nonlinear fits, we estimate the visibilities of the HOM dip to $58\pm10\%$ and of the HOM maximum to $34\pm10\%$ which is smaller than the expected visibilities (98\% and 50\%, respectively).This discrepancy can be due to an ensemble of causes:
\begin{itemize}
\item the asymmetry of the pump beam (Fig. \ref*{setup}c),
\item the asymmetry of the SPDC beams at the output of the crystal, because of the  walk-off in the crystal,
\item the cumulated geometric aberrations in the imaging systems and some residual misalignment,
\item some difference in the magnification of the optical systems in both arms.
\end{itemize}
 Of all the causes listed that may reduce the visibility of the HOM interference, geometric aberrations and residual misalignments are probably the main causes. These defects should be corrected by a better control of the alignment procedure. All these causes have effects on the spatial map of HOM Interference (see Fig. \ref{Joinproba}). Note, however, that this discrepancy cannot be due to the imperfect nature of detectors, whatever its nature, because these detectors are independent. The use of covariances, or correlations after normalization, eliminates the accidental coincidences: a positive covariance does correspond to a probability of detection of the photons of a pair on both detectors, i.e. imperfect HOM interference. 
 
 To conclude this part, we would like to emphasize that we have achieved the simultaneous two-photon interference of 1500 spatial modes (i.e. more than one order of magnitude more than in the paper of Lee et al.\cite{lee_spatial_2006}) with 60\% visibility. From the nonlinear fits of the experimental data depicted by the dotted curves in Fig. \ref{Diptemporel}, we also estimate the standard deviation of the gaussian shape of these curves to $\sigma_t=133.1\pm 0.2\,fs$ which corresponds to a wavelength standard deviation $\sigma_\lambda=\frac{\lambda^2_{SPDC}}{2\pi c\sigma_t}=2\,nm$ ($\lambda_{SPDC}=709\,nm$), in good agreement with the FWHM bandwidth of the interferential filter. \textcolor{blue}{\textbf{As EMCCD cameras have no temporal resolution, temporal entanglement cannot be measured directly. However, as the width of the temporal HOM dip is related to the coherence time of the biphoton state, we can estimate the temporal Schmidt number as $K_t=\frac{1}{2}\frac{\sigma_t^{pump}}{\sigma_t}=\frac{1}{2}\frac{400\times10^{-12}}{133\times10^{-15}}\simeq 1500$ which is of the same order of magnitude with the estimated temporal Schmidt number (Eq. \ref{eq2}).}}
     
\begin{figure}
  \centering
  \includegraphics[width=15cm]{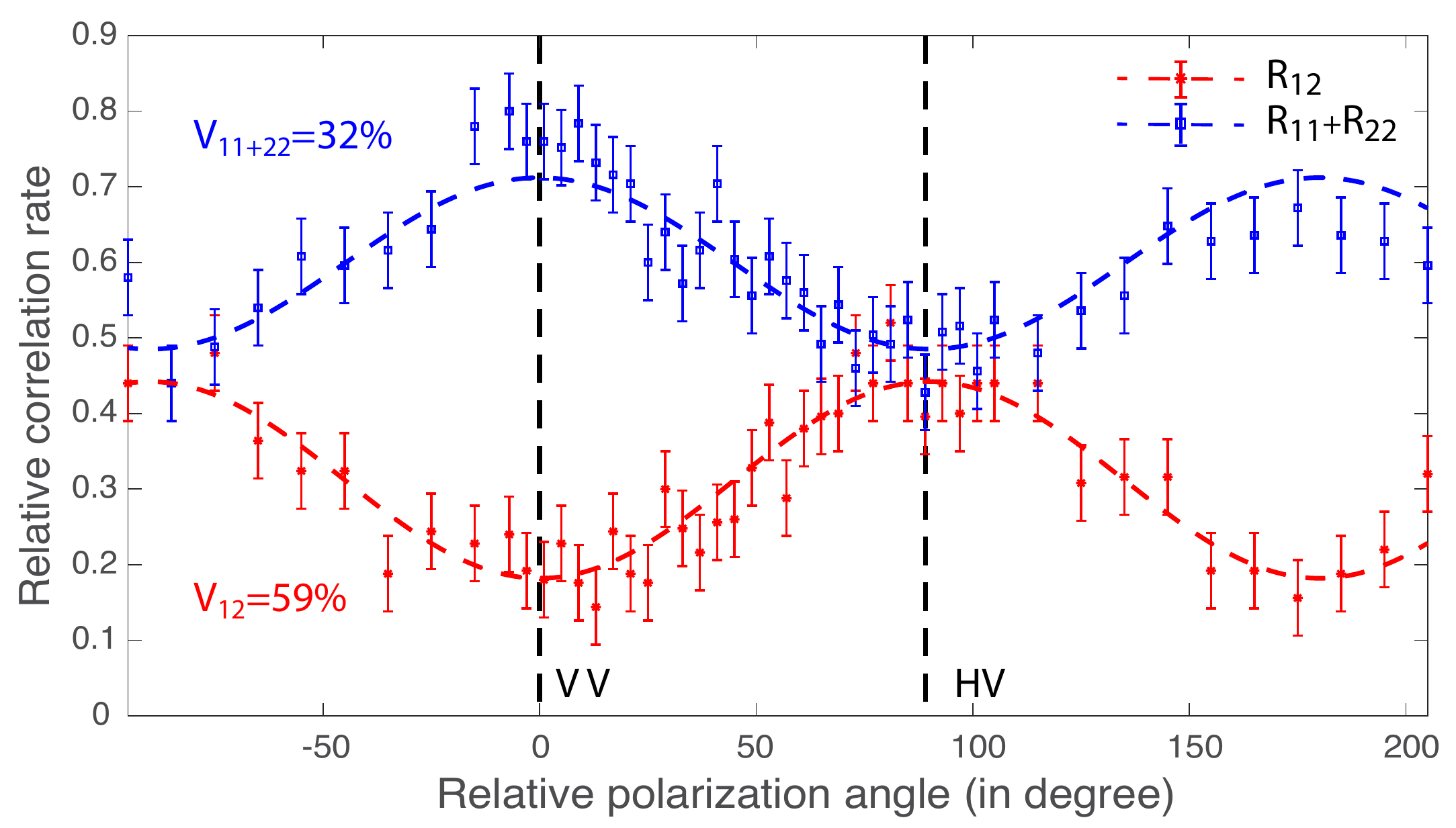}
  \caption{Correlation ratios as a function of the polarization angle between the two SPDC beams. Red stars and blue squares correspond respectively to correlations between the two images ($R_{12}$) and within single images ($R_{11}+R_{22}$). The dotted curves correspond to nonlinear fits of the experimental data.}\label{HOMPolar}
 \end{figure}

 The next experiment consists in measuring the variation of the relative coincidence ratios $R_{12}$ and $R_{11}+R_{22}$ as a function of the angle between the polarization directions of the twin beams, controlled by means of the HWP. These experimental results, depicted in Fig. \ref{HOMPolar}, clearly show the typical modulation of the HOM interference versus the relative polarization directions of the twin photons with visibilities $V_{12}=59\pm 10\%$ and $V_{11+22}=32\pm 10\%$, respectively. 
       
 \subsection{Spatial coherence measurements}
 
\begin{figure}
       \centering
       \includegraphics[width=15cm]{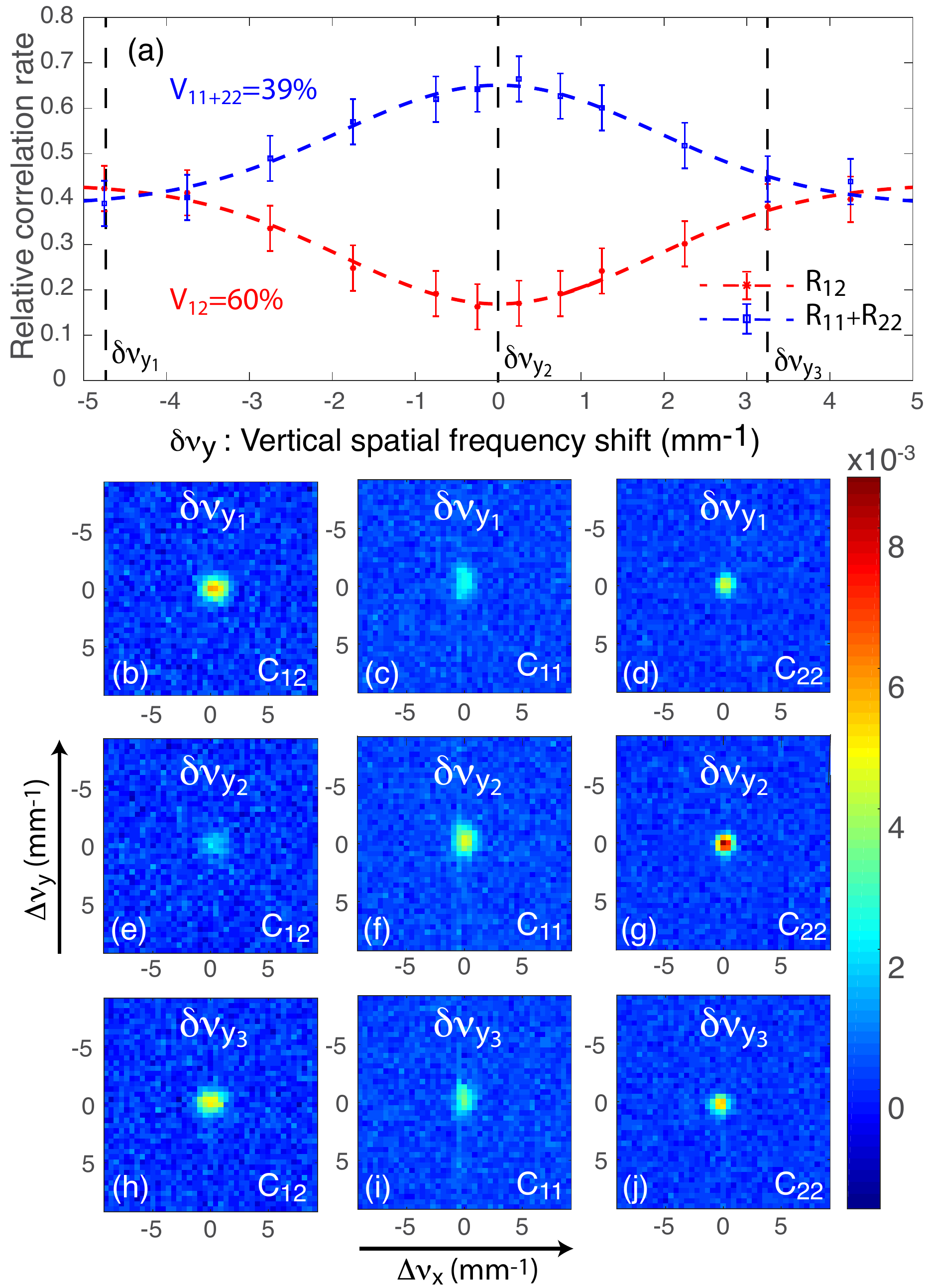}
       \caption{(a) Relative correlation ratios as a function of the vertical spatial frequency shift $\delta\nu_y$. (b-j) Average normalized spatial momentum correlations distributions $C_{12}$, $C_{11}$ and $C_{22}$ for different values of the vertical shift.}\label{DipspatialV}
      \end{figure}

\begin{figure} 
 \centering
 \includegraphics[width=15cm]{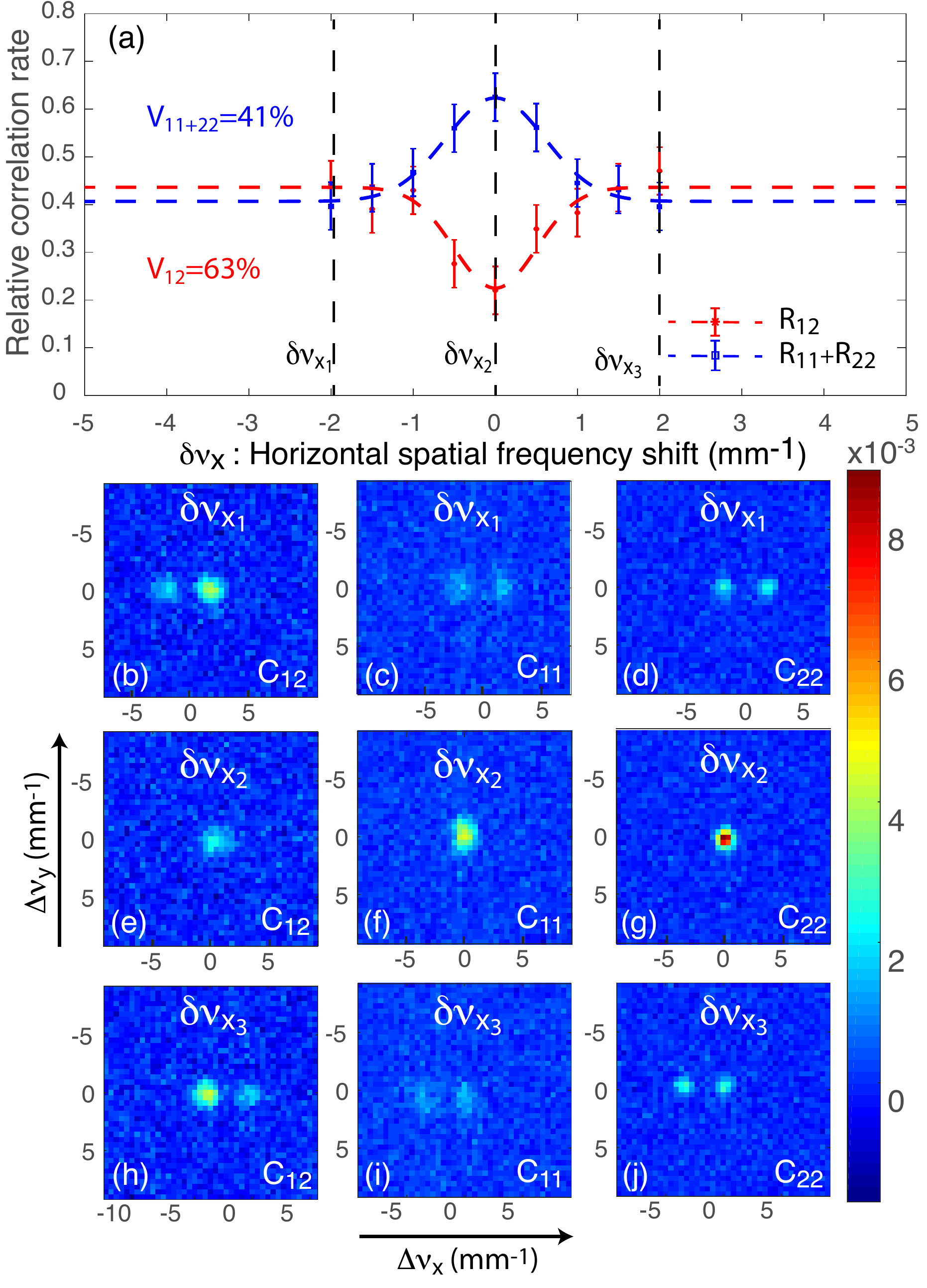}
 \caption{(a) Relative correlation ratios as a function of the horizontal spatial frequency shift $\delta\nu_x$. (b-j) Average normalized spatial momentum correlations distributions $C_{12}$, $C_{11}$ and $C_{22}$ for different values of the horizontal shift.}\label{DipspatialH}
  \end{figure}

 To measure the 2D spatial coherence of the biphoton state, the polarization states of the SPDC beams and the time delay are adjusted to obtain the best polarization and temporal indiscernability of the twin photons. From this initial configuration, vertical and horizontal shifts ($\delta\nu_{y}$, $\delta\nu_{x}$) between the reflected and transmitted beams are induced by tilting the BS around its horizontal and vertical axes, respectively (Fig. \ref{setup}). Then, the variation of the relative spatial correlations ratios between the twin images and within single images is measured as a function of the spatial frequency shifts. These experimental results are given in figures \ref{DipspatialV} and \ref{DipspatialH}.         
 
 In Fig. \ref{DipspatialV}, we present images of the 2D momentum correlations distributions $C_{12}(\Delta\boldsymbol{q})$, $C_{11}(\Delta\boldsymbol{q})$ and $C_{22}(\Delta\boldsymbol{q})$ for three different values of the vertical spatial frequency shift : $\delta\nu_y=$-4.75, 0 and +3.25 $mm^{-1}$ (Figures \ref{DipspatialV}b to \ref{DipspatialV}j). As it was predicted in \cite{devaux_stochastic_2019}, single spatial correlation peaks are observed and their positions do not depend on the vertical spatial shift of the reflected beam. Indeed, because of the geometry of the imaging system, vertical shifts of the reflected beams are in opposite directions on each camera (Fig. \ref{IntraInter}). Consequently, the spatial momentum correlations $C_{12}$ between transmitted-transmitted or reflected-reflected twin photons occur between opposite pixels of the twin images with no shift of the symmetry center. Similarly, the spatial momentum correlations $C_{11}$ and $C_{22}$ between transmitted-reflected or reflected-transmitted twin photons are measured between symmetric pixels along the vertical axis within single images of both cameras, also with the same symmetry center. In that case, the effect of the vertical shift only results in a variation of the relative total number of spatial coincidences between images and within single images (Fig. \ref{DipspatialV}a), where $R_{12}$ exhibits a spatial HOM dip and where $R_{11}+R_{22}$ exhibits a HOM maximum with  visibilities $V_{12}=60\pm 10\%$ and $V_{11+22}=39\pm 10\%$, respectively. From these curves, we estimate the standard deviation of the dip, versus a vertical frequency shift $\delta\nu_y$ induced by the rotation $\delta\phi_{BS}$ of the BS, to $2.7\pm0.3 mm^{-1}$, which is larger than the standard deviation of the correlation peaks in the far-field images along the vertical direction, i.e. versus the pixel coordinate $\Delta\nu_y$. According to the geometry of the HOM interferometer, the width of the spatial HOM dip along the vertical direction should be limited by the spatial phase matching bandwidth (Eq. \ref{eq4}) for perfect focusing. However, we have demonstrated in \cite{devaux_stochastic_2019} that a small defocusing between image planes reduces the vertical width of the HOM dip. As a perfect superposition of the image planes is experimentally difficult to achieve because of the geometric aberrations of the imaging systems, the measured standard deviation of the HOM dip along the vertical axis is, in agreement with our previous numerical results, smaller than the phase-matching bandwidth and greater than the coherence width of the biphoton state.
 
In Fig. \ref{DipspatialH}, we present the spatial momentum correlation distributions for three different values of the horizontal spatial frequency shift : $\delta\nu_x=\pm 2$ and $0\,mm^{-1}$ (Figures \ref{DipspatialH}b to \ref{DipspatialH}j). Contrary to the results for a vertical shift and in good agreement with \cite{devaux_stochastic_2019}, the spatial momentum correlations distributions exhibit two correlation peaks centered at $\pm\delta\nu_x$. In $C_{12}$, when $\delta\nu_x\neq 0$, the more intense peak corresponds to the $rr$ twin photons and the other peak corresponds to the $tt$ twin photons. Indeed, the amplitude difference between the two correlations peaks is due to the difference between the reflection and transmission coefficients ($R>T$). Moreover, the distance between the two correlation peaks corresponds to $2\delta\nu_x$. For $C_{11}$ and $C_{22}$, the two correlation peaks are of the same intensity because correlations are calculated between the upper and the lower parts of images where the transmitted and the reflected photons are equally distributed over the area of the detectors. Fig. \ref{DipspatialH}a shows the variation of the relative total number of spatial coincidences between images and within single images as a function of $\delta\nu_x$ where $R_{12}$ exhibits a spatial HOM dip and where $R_{11}+R_{22}$ exhibits a HOM maximum with visibilities $V_{12}=63\pm 10\%$ and $V_{11+22}=41\pm 10\%$, respectively. From these curves, we estimate the standard deviations of the dip to $0.7\pm0.1$ $mm^{-1}$, in good agreement with the standard deviation of the correlation peak along the horizontal direction.
Figures \ref{DipspatialV}a and \ref{DipspatialH}a show vertical and horizontal cross-sections of the 2D HOM dip and HOM maximum. Two-dimensional mapping of the HOM dip (like in figures 5a and 6 in \cite{devaux_stochastic_2019}) requires a high degree of accuracy for the spatial frequency scanning and stability of the setup during the necessarily long acquisition time. These conditions have not yet been fulfilled in our setup. Among the different improvements of the experimental setup to increase the visibility of the HOM interference, more accurate and automated scanning of the BS should allow obtaining a 2D map of the HOM dip with good resolution.

\begin{figure} 
  	\centering
  	\includegraphics[width=15cm]{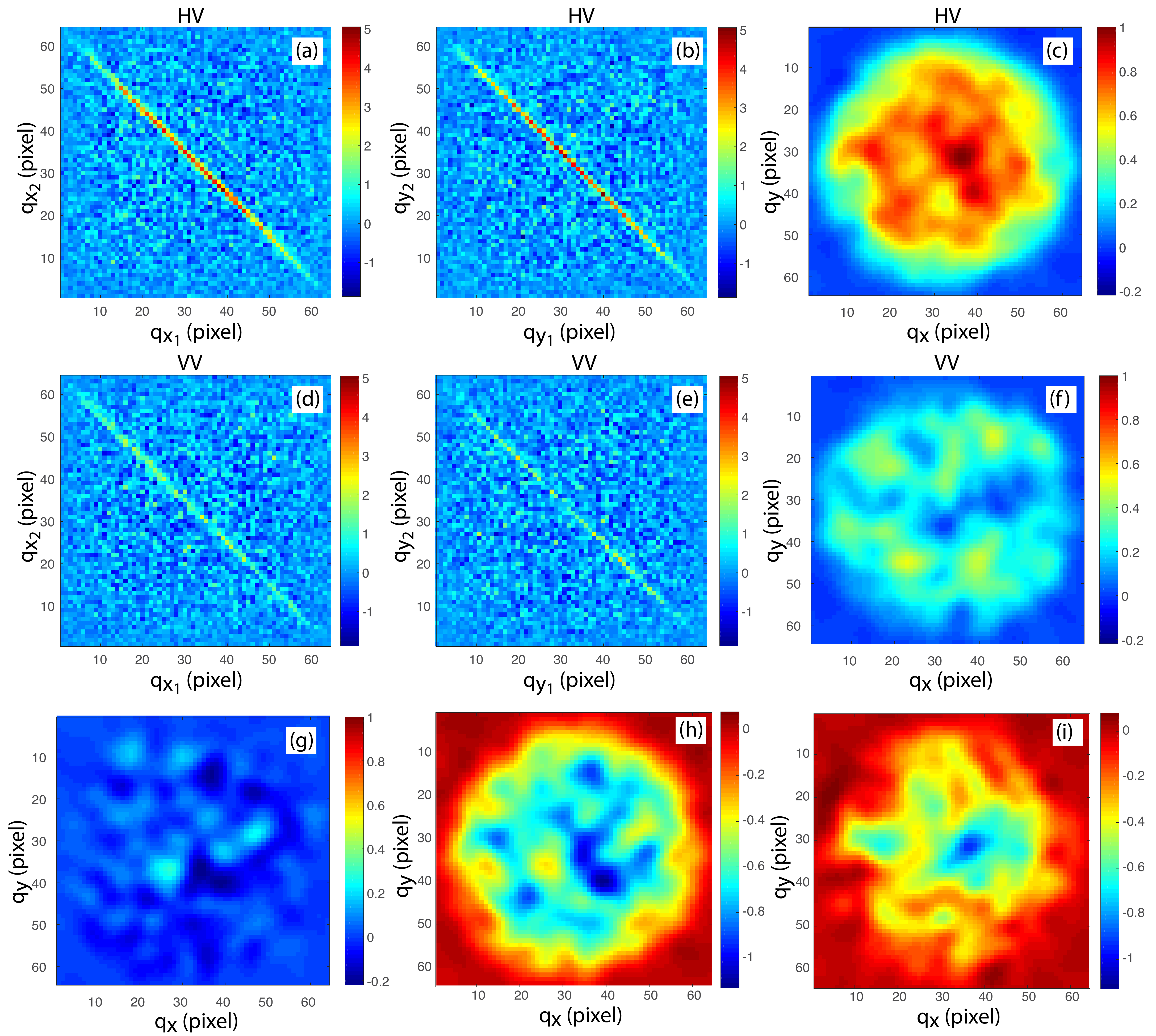}
  
  	\caption{(a) to (f) : Average covariance maps of the intensity fluctuations between pixels of 500 twin SPDC images versus 1D and 2D spatial transverse coordinates $q_x$ and $q_y$, for crossed (HV) or parallel (VV) polarizations. (g) Covariance map of the intensity fluctuations between pixels of 500 independant SPDC images versus 2D spatial transerve coordinates, normalized by the maximum covariance for HV polarization. (h)  Difference between the covariance maps (g) and (c). (i)  Difference between the covariance maps (f) and (c).}\label{Joinproba}
  \end{figure}

  Finally, we present in figures \ref*{Joinproba}a to \ref*{Joinproba}f the average correlation maps of the intensity fluctuations calculated between 4$\times$4 binned pixels of 500 twin SPDC images versus 1D and 2D spatial transverse coordinates $q_x$ and $q_y$ (given in pixels), for crossed ($HV$) and parallel ($VV$) polarizations when the twin beams are spatially and temporally superimposed ($\boldsymbol{\delta q}=\boldsymbol{0}$, $\delta t=0$, Fig. \ref{CORRspatialDiptemporel}). While the 1D maps give the classical representation of 1D far-field spatial correlations between twin photons, figures \ref*{Joinproba}c and \ref*{Joinproba}f show a first tentative of 2D resolution of the spatial coincidences between twin photons in the whole SPDC beams cross section. Fig.\ref*{Joinproba}g shows the same 2D map obtained with a set of two images issued from two different laser shots. Amplitudes of these maps are normalized by the maximum of the 2D correlation map corresponding to the HV configuration (Fig. \ref*{Joinproba}c). From these 2D maps, we have calculated the spatial distribution of the fall of coincidences between the $HV$ and the $VV$ maps (Fig. \ref*{Joinproba}i) and the fall obtained when correlating two images issued from two different laser shots (Fig. \ref*{Joinproba}h). We see clearly that the HOM results are close to ideal near the image center, but poorer near the edges. To obtain these images from 500 couple of experimental images, we had to remove the random noise by, first, binning the pixels 4$\times$4, second convolving the mean correlation image with a Gaussian kernel, whose integral corresponds to 44 binned pixels. Theses operations induce a strong loss of resolution but are necessary to obtain reliable maps with 500 image pairs. We present in the Appendix a calculation  which shows that the chosen resolution ensures a sufficient  signal-to noise ratio in the images of Fig. \ref*{Joinproba}.

\section{Conclusion}
We have reported the first experimental observation of fully spatio-temporal HOM interference of biphoton state of extremely high Schmidt number. Two-photon interference of 1500 spatial modes and more than $3\times10^6$ spatio-temporal modes is evidenced by measuring momentum spatial coincidences between pixels of far-field images of two SPDC beams propagating through a HOM interferometer. The output beams are detected with two separate detectors arrays operating in the photon-counting regime. The properties of HOM interference are investigated both in the time and space domains. We show that the two-photon interference exhibits temporal and 2D spatial HOM dips with an average visibility of 60\% and widths in good agreement with the spatio-temporal coherence properties of the biphoton state and the geometry of the HOM interferometer. This relatively low visibility, compared to the expected value, is probably due to geometric aberrations of the imaging system. Moreover, we also demonstrate that, using detectors arrays, 2D momentum spatial coincidences are resolved between the two output ports as well as within the two single ports images. This gives access to the rates and the 2D momentum correlations distributions of twin photons detected in pairs between the two cameras and on each camera. We also emphasize that the temporal coherence of the bi-photon state is measured with detectors that record spatial coincidences on the whole set of photons, without any prior selection of the photons in time and space coincidence. Given the critical role played by two-photon HOM interferences in most quantum information and quantum technology schemes, our demonstration that HOM interference can be obtained by manipulating a very high dimensional entangled state paves the way to very high dimensional quantum information schemes using space and time variables.
For example, the teleportation protocol of Bouwmeester et al. \cite{bouwmeester_experimental_1997} used a HOM set-up to perform the discrimination of the Bell states. Hence, our experiment paves the way to the teleportation of images, despite the fact that it can be proved \cite{calsamiglia_generalized_2002} that a purely linear system cannot be used to teleport a qudit, where the $d$ of qudit means a dimension greater than 2. Nevertheless, a teleportation of a qutrit has been recently reported \cite{luo_quantum_2019}, by using ancilla photons and a supplementary dimension. Further investigation is necessary to understand how to use mastering of spatial aspects of HOM interference in such type of protocol. Because some cryptography protocols use teleportation, applications to cryptography in multimodes fibers could follow. Another application concerns communications in quantum computing \cite{llewellyn_chip--chip_2020}.
\section*{Funding}
This work was partly supported by the French "Investissements d'Avenir" program, project ISITE-BFC (contract ANR-15-IDEX-03).

\section*{Appendix}
We calculate in this Appendix the signal-to-noise ratio obtained in the images of Fig. \ref*{Joinproba}, obtained, first,  by averaging 500 experimental images, second by binning the pixels $4\times 4$ and convolving the binned image by a Gaussian kernel. Hence, the resolution cell in the $256\times 256$ pixels  images of Fig. \ref*{Joinproba} contains $P=23\times 23$ pixels.\\
On each image, the thresholded number of photons, 0 or 1, follows a Bernouilli law of mean $m$ (here, $m=0.12$) and of variance $m(1-m)$.
For independent images the true covariance vanishes, but the estimator of covariance experiences fluctuations. If we use $N$ images and $P$ pixels in each image, which can be considered independent in this calculation, the variance $V$ of this estimator is  given by:
$V=\frac{m^2\times (1-m^2)}{NP}$. For $m=0.12$ photon/pixel, N=500 images, and $P=23\times23$ pixels,  we obtain a standard deviation $V^{1/2}=2.31\times 10^{-4}$ photon/pixel.

On the other hand, the twin signal $t$ can be estimated as $t=\frac{m}{P\prime}\times 13 \%=5.8\times {10^{-4}}$ photon/pixel, where 13\% is the ratio in $C_{12}$ given above and $P^\prime=27$ pixels the normalized integral of the correlation peak. By binning the pixels $4\times4$, this signal is, experimentally, multiplied by 10, because of the partial correlation between neighbor pixels, while the standard deviation of the covariance is only multiplied by 4 (see above, P divided by 16). This gives a signal-to-noise ratio of approximately 6 for the twin signal (Fig. \ref*{Joinproba}c) and of 3 for the HOM signal (Fig. \ref*{Joinproba} f). This is the minimum to observe a clear spatially deterministic signal in the HOM signal, that allows a first experimental map of its quality, but with a resolution that remains to be improved.

\bibliography{manipHOM.bib}

\begin{thebibliography}{32}
\expandafter\ifx\csname natexlab\endcsname\relax\def\natexlab#1{#1}\fi
\expandafter\ifx\csname bibnamefont\endcsname\relax
  \def\bibnamefont#1{#1}\fi
\expandafter\ifx\csname bibfnamefont\endcsname\relax
  \def\bibfnamefont#1{#1}\fi
\expandafter\ifx\csname citenamefont\endcsname\relax
  \def\citenamefont#1{#1}\fi
\expandafter\ifx\csname url\endcsname\relax
  \def\url#1{\texttt{#1}}\fi
\expandafter\ifx\csname urlprefix\endcsname\relax\def\urlprefix{URL }\fi
\providecommand{\bibinfo}[2]{#2}
\providecommand{\eprint}[2][]{\url{#2}}

\bibitem[{\citenamefont{Lubin et~al.}(2019)\citenamefont{Lubin, Tenne,
  Antolovic, Charbon, Bruschini, and Oron}}]{lubin_quantum_2019}
\bibinfo{author}{\bibfnamefont{G.}~\bibnamefont{Lubin}},
  \bibinfo{author}{\bibfnamefont{R.}~\bibnamefont{Tenne}},
  \bibinfo{author}{\bibfnamefont{I.~M.} \bibnamefont{Antolovic}},
  \bibinfo{author}{\bibfnamefont{E.}~\bibnamefont{Charbon}},
  \bibinfo{author}{\bibfnamefont{C.}~\bibnamefont{Bruschini}},
  \bibnamefont{and} \bibinfo{author}{\bibfnamefont{D.}~\bibnamefont{Oron}},
  \bibinfo{journal}{Opt. Express, OE} \textbf{\bibinfo{volume}{27}},
  \bibinfo{pages}{32863} (\bibinfo{year}{2019}), ISSN
  \bibinfo{issn}{1094-4087},
  \urlprefix\url{https://www.osapublishing.org/oe/abstract.cfm?uri=oe-27-23-32863}.

\bibitem[{\citenamefont{Moreau et~al.}(2019)\citenamefont{Moreau, Toninelli,
  Gregory, and Padgett}}]{moreau_imaging_2019}
\bibinfo{author}{\bibfnamefont{P.-A.} \bibnamefont{Moreau}},
  \bibinfo{author}{\bibfnamefont{E.}~\bibnamefont{Toninelli}},
  \bibinfo{author}{\bibfnamefont{T.}~\bibnamefont{Gregory}}, \bibnamefont{and}
  \bibinfo{author}{\bibfnamefont{M.~J.} \bibnamefont{Padgett}},
  \bibinfo{journal}{Nature Reviews Physics} p.~\bibinfo{pages}{1}
  (\bibinfo{year}{2019}), ISSN \bibinfo{issn}{2522-5820},
  \urlprefix\url{https://www.nature.com/articles/s42254-019-0056-0}.

\bibitem[{\citenamefont{Moreau et~al.}(2012)\citenamefont{Moreau,
  Mougin-Sisini, Devaux, and Lantz}}]{moreau_realization_2012}
\bibinfo{author}{\bibfnamefont{P.-A.} \bibnamefont{Moreau}},
  \bibinfo{author}{\bibfnamefont{J.}~\bibnamefont{Mougin-Sisini}},
  \bibinfo{author}{\bibfnamefont{F.}~\bibnamefont{Devaux}}, \bibnamefont{and}
  \bibinfo{author}{\bibfnamefont{E.}~\bibnamefont{Lantz}},
  \bibinfo{journal}{Phys. Rev. A} \textbf{\bibinfo{volume}{86}},
  \bibinfo{pages}{010101} (\bibinfo{year}{2012}), \bibinfo{note}{publisher:
  American Physical Society},
  \urlprefix\url{https://link.aps.org/doi/10.1103/PhysRevA.86.010101}.

\bibitem[{\citenamefont{Moreau et~al.}(2014)\citenamefont{Moreau, Devaux, and
  Lantz}}]{moreau_einstein-podolsky-rosen_2014}
\bibinfo{author}{\bibfnamefont{P.-A.} \bibnamefont{Moreau}},
  \bibinfo{author}{\bibfnamefont{F.}~\bibnamefont{Devaux}}, \bibnamefont{and}
  \bibinfo{author}{\bibfnamefont{E.}~\bibnamefont{Lantz}},
  \bibinfo{journal}{Phys. Rev. Lett.} \textbf{\bibinfo{volume}{113}},
  \bibinfo{pages}{160401} (\bibinfo{year}{2014}),
  \urlprefix\url{https://link.aps.org/doi/10.1103/PhysRevLett.113.160401}.

\bibitem[{\citenamefont{Lantz et~al.}(2015)\citenamefont{Lantz, Denis, Moreau,
  and Devaux}}]{lantz_einstein-podolsky-rosen_2015}
\bibinfo{author}{\bibfnamefont{E.}~\bibnamefont{Lantz}},
  \bibinfo{author}{\bibfnamefont{S.}~\bibnamefont{Denis}},
  \bibinfo{author}{\bibfnamefont{P.-A.} \bibnamefont{Moreau}},
  \bibnamefont{and} \bibinfo{author}{\bibfnamefont{F.}~\bibnamefont{Devaux}},
  \bibinfo{journal}{Opt. Express, OE} \textbf{\bibinfo{volume}{23}},
  \bibinfo{pages}{26472} (\bibinfo{year}{2015}), ISSN
  \bibinfo{issn}{1094-4087},
  \urlprefix\url{https://www.osapublishing.org/oe/abstract.cfm?uri=oe-23-20-26472}.

\bibitem[{\citenamefont{Edgar et~al.}(2012)\citenamefont{Edgar, Tasca,
  Izdebski, Warburton, Leach, Agnew, Buller, Boyd, and
  Padgett}}]{edgar_imaging_2012}
\bibinfo{author}{\bibfnamefont{M.~P.} \bibnamefont{Edgar}},
  \bibinfo{author}{\bibfnamefont{D.~S.} \bibnamefont{Tasca}},
  \bibinfo{author}{\bibfnamefont{F.}~\bibnamefont{Izdebski}},
  \bibinfo{author}{\bibfnamefont{R.~E.} \bibnamefont{Warburton}},
  \bibinfo{author}{\bibfnamefont{J.}~\bibnamefont{Leach}},
  \bibinfo{author}{\bibfnamefont{M.}~\bibnamefont{Agnew}},
  \bibinfo{author}{\bibfnamefont{G.~S.} \bibnamefont{Buller}},
  \bibinfo{author}{\bibfnamefont{R.~W.} \bibnamefont{Boyd}}, \bibnamefont{and}
  \bibinfo{author}{\bibfnamefont{M.~J.} \bibnamefont{Padgett}},
  \bibinfo{journal}{Nature Communications} \textbf{\bibinfo{volume}{3}},
  \bibinfo{pages}{984} (\bibinfo{year}{2012}), ISSN \bibinfo{issn}{2041-1723},
  \bibinfo{note}{number: 1 Publisher: Nature Publishing Group},
  \urlprefix\url{https://www.nature.com/articles/ncomms1988}.

\bibitem[{\citenamefont{Morris et~al.}(2015)\citenamefont{Morris, Aspden, Bell,
  Boyd, and Padgett}}]{morris_imaging_2015}
\bibinfo{author}{\bibfnamefont{P.~A.} \bibnamefont{Morris}},
  \bibinfo{author}{\bibfnamefont{R.~S.} \bibnamefont{Aspden}},
  \bibinfo{author}{\bibfnamefont{J.~E.~C.} \bibnamefont{Bell}},
  \bibinfo{author}{\bibfnamefont{R.~W.} \bibnamefont{Boyd}}, \bibnamefont{and}
  \bibinfo{author}{\bibfnamefont{M.~J.} \bibnamefont{Padgett}},
  \bibinfo{journal}{Nature Communications} \textbf{\bibinfo{volume}{6}},
  \bibinfo{pages}{5913} (\bibinfo{year}{2015}), ISSN \bibinfo{issn}{2041-1723},
  \urlprefix\url{https://www.nature.com/articles/ncomms6913}.

\bibitem[{\citenamefont{Denis et~al.}(2017)\citenamefont{Denis, Moreau, Devaux,
  and Lantz}}]{denis_temporal_2017}
\bibinfo{author}{\bibfnamefont{S.}~\bibnamefont{Denis}},
  \bibinfo{author}{\bibfnamefont{P.-A.} \bibnamefont{Moreau}},
  \bibinfo{author}{\bibfnamefont{F.}~\bibnamefont{Devaux}}, \bibnamefont{and}
  \bibinfo{author}{\bibfnamefont{E.}~\bibnamefont{Lantz}}, \bibinfo{journal}{J.
  Opt.} \textbf{\bibinfo{volume}{19}}, \bibinfo{pages}{034002}
  (\bibinfo{year}{2017}), ISSN \bibinfo{issn}{2040-8986},
  \urlprefix\url{http://stacks.iop.org/2040-8986/19/i=3/a=034002}.

\bibitem[{\citenamefont{Defienne et~al.}(2018)\citenamefont{Defienne, Reichert,
  and Fleischer}}]{defienne_adaptive_2018}
\bibinfo{author}{\bibfnamefont{H.}~\bibnamefont{Defienne}},
  \bibinfo{author}{\bibfnamefont{M.}~\bibnamefont{Reichert}}, \bibnamefont{and}
  \bibinfo{author}{\bibfnamefont{J.~W.} \bibnamefont{Fleischer}},
  \bibinfo{journal}{Phys. Rev. Lett.} \textbf{\bibinfo{volume}{121}},
  \bibinfo{pages}{233601} (\bibinfo{year}{2018}),
  \urlprefix\url{https://link.aps.org/doi/10.1103/PhysRevLett.121.233601}.

\bibitem[{\citenamefont{Devaux et~al.}(2019{\natexlab{a}})\citenamefont{Devaux,
  Mosset, Bassignot, and Lantz}}]{devaux_quantum_2019}
\bibinfo{author}{\bibfnamefont{F.}~\bibnamefont{Devaux}},
  \bibinfo{author}{\bibfnamefont{A.}~\bibnamefont{Mosset}},
  \bibinfo{author}{\bibfnamefont{F.}~\bibnamefont{Bassignot}},
  \bibnamefont{and} \bibinfo{author}{\bibfnamefont{E.}~\bibnamefont{Lantz}},
  \bibinfo{journal}{Phys. Rev. A} \textbf{\bibinfo{volume}{99}},
  \bibinfo{pages}{033854} (\bibinfo{year}{2019}{\natexlab{a}}),
  \urlprefix\url{https://link.aps.org/doi/10.1103/PhysRevA.99.033854}.

\bibitem[{\citenamefont{Brida et~al.}(2010)\citenamefont{Brida, Genovese, and
  Berchera}}]{brida_experimental_2010}
\bibinfo{author}{\bibfnamefont{G.}~\bibnamefont{Brida}},
  \bibinfo{author}{\bibfnamefont{M.}~\bibnamefont{Genovese}}, \bibnamefont{and}
  \bibinfo{author}{\bibfnamefont{I.~R.} \bibnamefont{Berchera}},
  \bibinfo{journal}{Nature Photonics} \textbf{\bibinfo{volume}{4}},
  \bibinfo{pages}{227} (\bibinfo{year}{2010}), ISSN \bibinfo{issn}{1749-4893},
  \urlprefix\url{https://www.nature.com/articles/nphoton.2010.29}.

\bibitem[{\citenamefont{Toninelli et~al.}(2017)\citenamefont{Toninelli, Edgar,
  Moreau, Gibson, Hammond, and Padgett}}]{toninelli_sub-shot-noise_2017}
\bibinfo{author}{\bibfnamefont{E.}~\bibnamefont{Toninelli}},
  \bibinfo{author}{\bibfnamefont{M.~P.} \bibnamefont{Edgar}},
  \bibinfo{author}{\bibfnamefont{P.-A.} \bibnamefont{Moreau}},
  \bibinfo{author}{\bibfnamefont{G.~M.} \bibnamefont{Gibson}},
  \bibinfo{author}{\bibfnamefont{G.~D.} \bibnamefont{Hammond}},
  \bibnamefont{and} \bibinfo{author}{\bibfnamefont{M.~J.}
  \bibnamefont{Padgett}}, \bibinfo{journal}{Opt. Express, OE}
  \textbf{\bibinfo{volume}{25}}, \bibinfo{pages}{21826} (\bibinfo{year}{2017}),
  ISSN \bibinfo{issn}{1094-4087},
  \urlprefix\url{https://www.osapublishing.org/oe/abstract.cfm?uri=oe-25-18-21826}.

\bibitem[{\citenamefont{Lemos et~al.}(2014)\citenamefont{Lemos, Borish, Cole,
  Ramelow, Lapkiewicz, and Zeilinger}}]{lemos_quantum_2014}
\bibinfo{author}{\bibfnamefont{G.~B.} \bibnamefont{Lemos}},
  \bibinfo{author}{\bibfnamefont{V.}~\bibnamefont{Borish}},
  \bibinfo{author}{\bibfnamefont{G.~D.} \bibnamefont{Cole}},
  \bibinfo{author}{\bibfnamefont{S.}~\bibnamefont{Ramelow}},
  \bibinfo{author}{\bibfnamefont{R.}~\bibnamefont{Lapkiewicz}},
  \bibnamefont{and}
  \bibinfo{author}{\bibfnamefont{A.}~\bibnamefont{Zeilinger}},
  \bibinfo{journal}{Nature} \textbf{\bibinfo{volume}{512}},
  \bibinfo{pages}{409} (\bibinfo{year}{2014}), ISSN \bibinfo{issn}{1476-4687},
  \urlprefix\url{https://www.nature.com/articles/nature13586}.

\bibitem[{\citenamefont{Hong et~al.}(1987)\citenamefont{Hong, Ou, and
  Mandel}}]{hong_measurement_1987}
\bibinfo{author}{\bibfnamefont{C.~K.} \bibnamefont{Hong}},
  \bibinfo{author}{\bibfnamefont{Z.~Y.} \bibnamefont{Ou}}, \bibnamefont{and}
  \bibinfo{author}{\bibfnamefont{L.}~\bibnamefont{Mandel}},
  \bibinfo{journal}{Phys. Rev. Lett.} \textbf{\bibinfo{volume}{59}},
  \bibinfo{pages}{2044} (\bibinfo{year}{1987}),
  \urlprefix\url{https://link.aps.org/doi/10.1103/PhysRevLett.59.2044}.

\bibitem[{\citenamefont{Simon et~al.}(2017)\citenamefont{Simon, Jaeger, and
  Sergienko}}]{simon_quantum_2017}
\bibinfo{author}{\bibfnamefont{D.~S.} \bibnamefont{Simon}},
  \bibinfo{author}{\bibfnamefont{G.}~\bibnamefont{Jaeger}}, \bibnamefont{and}
  \bibinfo{author}{\bibfnamefont{A.~V.} \bibnamefont{Sergienko}},
  \emph{\bibinfo{title}{Quantum {Metrology}, {Imaging}, and {Communication}}},
  Quantum {Science} and {Technology} (\bibinfo{publisher}{Springer
  International Publishing}, \bibinfo{year}{2017}), ISBN
  \bibinfo{isbn}{978-3-319-46549-4},
  \urlprefix\url{//www.springer.com/it/book/9783319465494}.

\bibitem[{\citenamefont{Bouwmeester et~al.}(1997)\citenamefont{Bouwmeester,
  Pan, Mattle, Eibl, Weinfurter, and
  Zeilinger}}]{bouwmeester_experimental_1997}
\bibinfo{author}{\bibfnamefont{D.}~\bibnamefont{Bouwmeester}},
  \bibinfo{author}{\bibfnamefont{J.-W.} \bibnamefont{Pan}},
  \bibinfo{author}{\bibfnamefont{K.}~\bibnamefont{Mattle}},
  \bibinfo{author}{\bibfnamefont{M.}~\bibnamefont{Eibl}},
  \bibinfo{author}{\bibfnamefont{H.}~\bibnamefont{Weinfurter}},
  \bibnamefont{and}
  \bibinfo{author}{\bibfnamefont{A.}~\bibnamefont{Zeilinger}},
  \bibinfo{journal}{Nature} \textbf{\bibinfo{volume}{390}},
  \bibinfo{pages}{575} (\bibinfo{year}{1997}), ISSN \bibinfo{issn}{1476-4687},
  \urlprefix\url{https://www.nature.com/articles/37539}.

\bibitem[{\citenamefont{Kok et~al.}(2007)\citenamefont{Kok, Munro, Nemoto,
  Ralph, Dowling, and Milburn}}]{kok_linear_2007}
\bibinfo{author}{\bibfnamefont{P.}~\bibnamefont{Kok}},
  \bibinfo{author}{\bibfnamefont{W.~J.} \bibnamefont{Munro}},
  \bibinfo{author}{\bibfnamefont{K.}~\bibnamefont{Nemoto}},
  \bibinfo{author}{\bibfnamefont{T.~C.} \bibnamefont{Ralph}},
  \bibinfo{author}{\bibfnamefont{J.~P.} \bibnamefont{Dowling}},
  \bibnamefont{and} \bibinfo{author}{\bibfnamefont{G.~J.}
  \bibnamefont{Milburn}}, \bibinfo{journal}{Rev. Mod. Phys.}
  \textbf{\bibinfo{volume}{79}}, \bibinfo{pages}{135} (\bibinfo{year}{2007}),
  \urlprefix\url{https://link.aps.org/doi/10.1103/RevModPhys.79.135}.

\bibitem[{\citenamefont{Gard et~al.}(2015)\citenamefont{Gard, Motes, Olson,
  Rohde, and Dowling}}]{gard_introduction_2015}
\bibinfo{author}{\bibfnamefont{B.~T.} \bibnamefont{Gard}},
  \bibinfo{author}{\bibfnamefont{K.~R.} \bibnamefont{Motes}},
  \bibinfo{author}{\bibfnamefont{J.~P.} \bibnamefont{Olson}},
  \bibinfo{author}{\bibfnamefont{P.~P.} \bibnamefont{Rohde}}, \bibnamefont{and}
  \bibinfo{author}{\bibfnamefont{J.~P.} \bibnamefont{Dowling}}, in
  \emph{\bibinfo{booktitle}{From {Atomic} to {Mesoscale}}}
  (\bibinfo{publisher}{WORLD SCIENTIFIC}, \bibinfo{year}{2015}), pp.
  \bibinfo{pages}{167--192}, ISBN \bibinfo{isbn}{978-981-4678-69-8},
  \urlprefix\url{https://www.worldscientific.com/doi/abs/10.1142/9789814678704_0008}.

\bibitem[{\citenamefont{Deng et~al.}(2019)\citenamefont{Deng, Wang, Ding, Duan,
  Qin, Chen, He, He, Li, Li et~al.}}]{deng_quantum_2019}
\bibinfo{author}{\bibfnamefont{Y.-H.} \bibnamefont{Deng}},
  \bibinfo{author}{\bibfnamefont{H.}~\bibnamefont{Wang}},
  \bibinfo{author}{\bibfnamefont{X.}~\bibnamefont{Ding}},
  \bibinfo{author}{\bibfnamefont{Z.-C.} \bibnamefont{Duan}},
  \bibinfo{author}{\bibfnamefont{J.}~\bibnamefont{Qin}},
  \bibinfo{author}{\bibfnamefont{M.-C.} \bibnamefont{Chen}},
  \bibinfo{author}{\bibfnamefont{Y.}~\bibnamefont{He}},
  \bibinfo{author}{\bibfnamefont{Y.-M.} \bibnamefont{He}},
  \bibinfo{author}{\bibfnamefont{J.-P.} \bibnamefont{Li}},
  \bibinfo{author}{\bibfnamefont{Y.-H.} \bibnamefont{Li}},
  \bibnamefont{et~al.}, \bibinfo{journal}{Phys. Rev. Lett.}
  \textbf{\bibinfo{volume}{123}}, \bibinfo{pages}{080401}
  (\bibinfo{year}{2019}),
  \urlprefix\url{https://link.aps.org/doi/10.1103/PhysRevLett.123.080401}.

\bibitem[{\citenamefont{Lee and van Exter}(2006)}]{lee_spatial_2006}
\bibinfo{author}{\bibfnamefont{P.~S.~K.} \bibnamefont{Lee}} \bibnamefont{and}
  \bibinfo{author}{\bibfnamefont{M.~P.} \bibnamefont{van Exter}},
  \bibinfo{journal}{Phys. Rev. A} \textbf{\bibinfo{volume}{73}},
  \bibinfo{pages}{063827} (\bibinfo{year}{2006}),
  \urlprefix\url{https://link.aps.org/doi/10.1103/PhysRevA.73.063827}.

\bibitem[{\citenamefont{Jachura and
  Chrapkiewicz}(2015)}]{jachura_shot-by-shot_2015}
\bibinfo{author}{\bibfnamefont{M.}~\bibnamefont{Jachura}} \bibnamefont{and}
  \bibinfo{author}{\bibfnamefont{R.}~\bibnamefont{Chrapkiewicz}},
  \bibinfo{journal}{Opt. Lett., OL} \textbf{\bibinfo{volume}{40}},
  \bibinfo{pages}{1540} (\bibinfo{year}{2015}), ISSN \bibinfo{issn}{1539-4794},
  \urlprefix\url{https://www.osapublishing.org/ol/abstract.cfm?uri=ol-40-7-1540}.

\bibitem[{\citenamefont{Chrapkiewicz et~al.}(2016)\citenamefont{Chrapkiewicz,
  Jachura, Banaszek, and Wasilewski}}]{chrapkiewicz_hologram_2016}
\bibinfo{author}{\bibfnamefont{R.}~\bibnamefont{Chrapkiewicz}},
  \bibinfo{author}{\bibfnamefont{M.}~\bibnamefont{Jachura}},
  \bibinfo{author}{\bibfnamefont{K.}~\bibnamefont{Banaszek}}, \bibnamefont{and}
  \bibinfo{author}{\bibfnamefont{W.}~\bibnamefont{Wasilewski}},
  \bibinfo{journal}{Nature Photon} \textbf{\bibinfo{volume}{10}},
  \bibinfo{pages}{576} (\bibinfo{year}{2016}), ISSN \bibinfo{issn}{1749-4893},
  \urlprefix\url{https://www.nature.com/articles/nphoton.2016.129}.

\bibitem[{\citenamefont{Ou and Mandel}(1989)}]{ou_further_1989}
\bibinfo{author}{\bibfnamefont{Z.~Y.} \bibnamefont{Ou}} \bibnamefont{and}
  \bibinfo{author}{\bibfnamefont{L.}~\bibnamefont{Mandel}},
  \bibinfo{journal}{Phys. Rev. Lett.} \textbf{\bibinfo{volume}{62}},
  \bibinfo{pages}{2941} (\bibinfo{year}{1989}),
  \urlprefix\url{https://link.aps.org/doi/10.1103/PhysRevLett.62.2941}.

\bibitem[{\citenamefont{Kim et~al.}(2006)\citenamefont{Kim, Kwon, Kim, and
  Kim}}]{kim_spatial_2006}
\bibinfo{author}{\bibfnamefont{H.}~\bibnamefont{Kim}},
  \bibinfo{author}{\bibfnamefont{O.}~\bibnamefont{Kwon}},
  \bibinfo{author}{\bibfnamefont{W.}~\bibnamefont{Kim}}, \bibnamefont{and}
  \bibinfo{author}{\bibfnamefont{T.}~\bibnamefont{Kim}},
  \bibinfo{journal}{Phys. Rev. A} \textbf{\bibinfo{volume}{73}},
  \bibinfo{pages}{023820} (\bibinfo{year}{2006}),
  \urlprefix\url{https://link.aps.org/doi/10.1103/PhysRevA.73.023820}.

\bibitem[{\citenamefont{Walborn et~al.}(2003)\citenamefont{Walborn,
  de~Oliveira, Pádua, and Monken}}]{walborn_multimode_2003}
\bibinfo{author}{\bibfnamefont{S.~P.} \bibnamefont{Walborn}},
  \bibinfo{author}{\bibfnamefont{A.~N.} \bibnamefont{de~Oliveira}},
  \bibinfo{author}{\bibfnamefont{S.}~\bibnamefont{Pádua}}, \bibnamefont{and}
  \bibinfo{author}{\bibfnamefont{C.~H.} \bibnamefont{Monken}},
  \bibinfo{journal}{Phys. Rev. Lett.} \textbf{\bibinfo{volume}{90}},
  \bibinfo{pages}{143601} (\bibinfo{year}{2003}),
  \urlprefix\url{https://link.aps.org/doi/10.1103/PhysRevLett.90.143601}.

\bibitem[{\citenamefont{Di~Lorenzo~Pires
  et~al.}(2010)\citenamefont{Di~Lorenzo~Pires, Florijn, and van
  Exter}}]{di_lorenzo_pires_measurement_2010}
\bibinfo{author}{\bibfnamefont{H.}~\bibnamefont{Di~Lorenzo~Pires}},
  \bibinfo{author}{\bibfnamefont{H.~C.~B.} \bibnamefont{Florijn}},
  \bibnamefont{and} \bibinfo{author}{\bibfnamefont{M.~P.} \bibnamefont{van
  Exter}}, \bibinfo{journal}{Phys. Rev. Lett.} \textbf{\bibinfo{volume}{104}},
  \bibinfo{pages}{020505} (\bibinfo{year}{2010}), \bibinfo{note}{publisher:
  American Physical Society},
  \urlprefix\url{https://link.aps.org/doi/10.1103/PhysRevLett.104.020505}.

\bibitem[{\citenamefont{Devaux et~al.}(2019{\natexlab{b}})\citenamefont{Devaux,
  Mosset, and Lantz}}]{devaux_stochastic_2019}
\bibinfo{author}{\bibfnamefont{F.}~\bibnamefont{Devaux}},
  \bibinfo{author}{\bibfnamefont{A.}~\bibnamefont{Mosset}}, \bibnamefont{and}
  \bibinfo{author}{\bibfnamefont{E.}~\bibnamefont{Lantz}},
  \bibinfo{journal}{Phys. Rev. A} \textbf{\bibinfo{volume}{100}},
  \bibinfo{pages}{013845} (\bibinfo{year}{2019}{\natexlab{b}}),
  \urlprefix\url{https://link.aps.org/doi/10.1103/PhysRevA.100.013845}.

\bibitem[{\citenamefont{Lantz et~al.}(2008)\citenamefont{Lantz, Blanchet,
  Furfaro, and Devaux}}]{lantz_multi-imaging_2008}
\bibinfo{author}{\bibfnamefont{E.}~\bibnamefont{Lantz}},
  \bibinfo{author}{\bibfnamefont{J.-L.} \bibnamefont{Blanchet}},
  \bibinfo{author}{\bibfnamefont{L.}~\bibnamefont{Furfaro}}, \bibnamefont{and}
  \bibinfo{author}{\bibfnamefont{F.}~\bibnamefont{Devaux}},
  \bibinfo{journal}{Mon Not R Astron Soc} \textbf{\bibinfo{volume}{386}},
  \bibinfo{pages}{2262} (\bibinfo{year}{2008}), ISSN \bibinfo{issn}{0035-8711},
  \urlprefix\url{https://academic.oup.com/mnras/article/386/4/2262/1466802}.

\bibitem[{\citenamefont{Law and Eberly}(2004)}]{law_analysis_2004}
\bibinfo{author}{\bibfnamefont{C.~K.} \bibnamefont{Law}} \bibnamefont{and}
  \bibinfo{author}{\bibfnamefont{J.~H.} \bibnamefont{Eberly}},
  \bibinfo{journal}{Phys. Rev. Lett.} \textbf{\bibinfo{volume}{92}},
  \bibinfo{pages}{127903} (\bibinfo{year}{2004}),
  \urlprefix\url{https://link.aps.org/doi/10.1103/PhysRevLett.92.127903}.

\bibitem[{\citenamefont{Calsamiglia}(2002)}]{calsamiglia_generalized_2002}
\bibinfo{author}{\bibfnamefont{J.}~\bibnamefont{Calsamiglia}},
  \bibinfo{journal}{Phys. Rev. A} \textbf{\bibinfo{volume}{65}},
  \bibinfo{pages}{030301} (\bibinfo{year}{2002}), \bibinfo{note}{publisher:
  American Physical Society},
  \urlprefix\url{https://link.aps.org/doi/10.1103/PhysRevA.65.030301}.

\bibitem[{\citenamefont{Luo et~al.}(2019)\citenamefont{Luo, Zhong, Erhard,
  Wang, Peng, Krenn, Jiang, Li, Liu, Lu et~al.}}]{luo_quantum_2019}
\bibinfo{author}{\bibfnamefont{Y.-H.} \bibnamefont{Luo}},
  \bibinfo{author}{\bibfnamefont{H.-S.} \bibnamefont{Zhong}},
  \bibinfo{author}{\bibfnamefont{M.}~\bibnamefont{Erhard}},
  \bibinfo{author}{\bibfnamefont{X.-L.} \bibnamefont{Wang}},
  \bibinfo{author}{\bibfnamefont{L.-C.} \bibnamefont{Peng}},
  \bibinfo{author}{\bibfnamefont{M.}~\bibnamefont{Krenn}},
  \bibinfo{author}{\bibfnamefont{X.}~\bibnamefont{Jiang}},
  \bibinfo{author}{\bibfnamefont{L.}~\bibnamefont{Li}},
  \bibinfo{author}{\bibfnamefont{N.-L.} \bibnamefont{Liu}},
  \bibinfo{author}{\bibfnamefont{C.-Y.} \bibnamefont{Lu}},
  \bibnamefont{et~al.}, \bibinfo{journal}{Phys. Rev. Lett.}
  \textbf{\bibinfo{volume}{123}}, \bibinfo{pages}{070505}
  (\bibinfo{year}{2019}), \bibinfo{note}{publisher: American Physical Society},
  \urlprefix\url{https://link.aps.org/doi/10.1103/PhysRevLett.123.070505}.

\bibitem[{\citenamefont{Llewellyn et~al.}(2020)\citenamefont{Llewellyn, Ding,
  Faruque, Paesani, Bacco, Santagati, Qian, Li, Xiao, Huber
  et~al.}}]{llewellyn_chip--chip_2020}
\bibinfo{author}{\bibfnamefont{D.}~\bibnamefont{Llewellyn}},
  \bibinfo{author}{\bibfnamefont{Y.}~\bibnamefont{Ding}},
  \bibinfo{author}{\bibfnamefont{I.~I.} \bibnamefont{Faruque}},
  \bibinfo{author}{\bibfnamefont{S.}~\bibnamefont{Paesani}},
  \bibinfo{author}{\bibfnamefont{D.}~\bibnamefont{Bacco}},
  \bibinfo{author}{\bibfnamefont{R.}~\bibnamefont{Santagati}},
  \bibinfo{author}{\bibfnamefont{Y.-J.} \bibnamefont{Qian}},
  \bibinfo{author}{\bibfnamefont{Y.}~\bibnamefont{Li}},
  \bibinfo{author}{\bibfnamefont{Y.-F.} \bibnamefont{Xiao}},
  \bibinfo{author}{\bibfnamefont{M.}~\bibnamefont{Huber}},
  \bibnamefont{et~al.}, \bibinfo{journal}{Nat. Phys.}
  \textbf{\bibinfo{volume}{16}}, \bibinfo{pages}{148} (\bibinfo{year}{2020}),
  ISSN \bibinfo{issn}{1745-2481}, \bibinfo{note}{number: 2 Publisher: Nature
  Publishing Group},
  \urlprefix\url{https://www.nature.com/articles/s41567-019-0727-x}.

\end{thebibliography}

\end{document}